\documentclass[english,conference,final,nameinlink,noabbrev]{IEEEtran}

\usepackage{cite} 
\usepackage{balance}

\usepackage[utf8]{inputenc}
\usepackage[T1]{fontenc}
\usepackage{microtype}

\usepackage{babel}
\usepackage{csquotes}
\MakeOuterQuote{"}

\usepackage{lmodern}
\usepackage{MnSymbol}

\usepackage[svgnames]{xcolor}
\usepackage[subject={TODO}]{pdfcomment}

\usepackage{ctable}
\newcolumntype{R}{>{\raggedleft \arraybackslash}X}

\usepackage{listings}

\newcommand{\lineref}[2]{\hyperref[#1]{line~\ref*{#1:#2}}}
\newcommand{\linerefn}[2]{\hyperref[#1]{line~#2}}
\newcommand{\linesrefn}[2]{\hyperref[#1]{lines~#2}}
\usepackage{accsupp}
\newcommand\emptyaccsupp[1]{\BeginAccSupp{ActualText={}}#1\EndAccSupp{}}

\usepackage{hyperref}
\usepackage[all]{hypcap}
\usepackage{cleveref}

\newcommand{\lstnumberstyle}[1]{\tiny\emptyaccsupp{#1}}
\lstdefinestyle{BWStyle}{
	keywordstyle=\bfseries,
	stringstyle=\color{DimGray},
	commentstyle=\textsl,
}
\lstdefinelanguage{algorithm}{
	keywords={function, for, do, if, then, else, return, in_, is_a, or, and},
	morecomment=[l]{'},
	morecomment=[s]{/*}{*/},
	morestring=[b]",  
	sensitive=true,
}
\lstdefinelanguage{HanaSQL}[]{SQL}{
	morekeywords={replace,string,if,is,daysbetween,secondsbetween,weekday,adddays,addseconds,double, procedure,begin,declare,inout,call,return,returns},
	moredelim=**[is][\slshape]{^}{^},
	moredelim=**[is][\bfseries]{§}{§},
}
\lstdefinelanguage{Inline}{
	moredelim=**[is][\slshape]{^}{^},
	moredelim=**[is][\bfseries]{§}{§},
}

\lstMakeShortInline[basicstyle=\ttfamily,language={Inline},breaklines=true]`

\usepackage{listingsnumbermargin}
\newcommand{\tool}{TAR\-DISP}
\newcommand{\SQLextension}{Back-in-time SQL}

\hypersetup{
	pdftitle={Bringing Back-in-Time Debugging Down to the Database},
	pdfauthor={Arian Treffer, Michael Perscheid, Matthias Uflacker}
}

\begin{document}

\title{Bringing Back-in-Time Debugging\\ Down to the Database}

\author{\IEEEauthorblockN{Arian Treffer}
\IEEEauthorblockA{Hasso-Plattner-Institut\\
Potsdam, Germany\\
arian.treffer@hpi.de}
\and
\IEEEauthorblockN{Michael Perscheid}
\IEEEauthorblockA{SAP Innovation Center\\
Potsdam, Germany\\
michael.perscheid@sap.com}
\and
\IEEEauthorblockN{Matthias Uflacker}
\IEEEauthorblockA{Hasso-Plattner-Institut\\
Potsdam, Germany\\
matthias.uflacker@hpi.de}}



\maketitle
\begin{abstract}
With back-in-time debuggers, developers can explore what happened before observable failures by following infection chains back to their root causes.
While there are several such debuggers for object-oriented programming languages, we do not know of any back-in-time capabilities at the database-level.
Thus, if failures are caused by SQL scripts or stored procedures, developers have difficulties in understanding their unexpected behavior.

In this paper, we present an approach for bringing back-in-time debugging down to the SAP HANA in-memory database.
Our \tool\ debugger allows developers to step queries backwards and inspecting the database at previous and arbitrary points in time.
With the help of a SQL extension, we can express queries covering a period of execution time within a debugging session and handle large amounts of data with low overhead on performance and memory.
The entire approach has been evaluated within a development project at SAP and shows promising results with respect to the gathered developer feedback.
\end{abstract}



\section{Introduction}

Finding defects in code is a frequent task for every programmer and is often difficult even with a deep understanding of the system.
To localize failure causes, they examine involved program entities and distinguish relevant from irrelevant behavior and clean from infected state. 
However, common symbolic debuggers do not support identification of such infection chains very well because they only provide access to the last point of execution without access to the program history.
Back-in-time also known as omniscient debuggers simplify the debugging process by making it easier to follow cause-effect chains from the observable failure back to the causing defect~\cite{lewis_debugging_2003}.
These tools provide full access to past events so that developers can directly experiment with the entire infection chain. 

Even though back-in-time debuggers exist for many object-oriented programming languages~\cite{feldman_igor:_1988,lewis_debugging_2003}, there are none, to the best of our knowledge, that run on databases and support SQL or SQLScript\footnote{SQLScript is a proprietary SQL extension for stored procedures in SAP HANA~\cite{sqlScript}.}.
This is mainly because of two reasons. 
First, back-in-time debuggers typically create a significant overhead on performance and memory consumption~\cite{lewis_debugging_2003,pothier_scalable_2007,lienhard_practical_2008}.
It seems unfeasible to use a back-in-time debugger on top of a database script that processes billions of records.
Second, current back-in-time debugging concepts cannot handle side-effects outside their system that usually happen in writing operations during INSERT and UPDATE statements. 

Due to high performance requirements of handling big data in business applications, SAP has not only a strong demand to move code closer to data~\cite{plattner2015memory} but also the need to improve development tools at the database-level. 
As existing tools mostly work on the application level, are limited to specific points in time, or work only on an abstract query plan, developers are often left alone when it comes to debug and understand the results of their SQL and SQLScripts.
We argue that a back-in-time debugger at the bottom of the technology stack is able to close this gap and can support developers in developing and maintaining their queries more efficiently. 

In this paper, we bring the concept of back-in-time debugging to the database and present \emph{\tool} as an implementation of our approach.
The contributions of this paper are as follows:
\begin{itemize}
	\item \emph{\tool} ("Tracing And Recorded Data In Stored Procedures") is a back-in-time debugger for stored procedures which can be installed in the SAP HANA in-memory database and programming platform.
		Using \tool, developers can move freely through the execution time of a stored procedure and inspect control flow, variables, and intermediate results.
	
	\item \emph{\SQLextension} is an extension to SQL allowing developers to submit arbitrary queries against previous states of the database 
		and to compare multiple points in time with one query.
		\tool\ provides a console for developers to submit \SQLextension\ queries which use variables or points in time from the current debug session.

	\item \emph{Very low overhead} when recording run-time data and the efficient querying of past database states allow developers to use \tool\ as the default tool for debugging database scripts even on large data sets.
	
\end{itemize}

We evaluated \tool\ with the help of SAP colleagues who worked on a project called \emph{Point of Sales Explorer}~\cite{plattner2015memory}, which makes heavy use of SQLScript. 
The interviews indicated that failures in stored procedures can be investigated more efficiently with \tool\ than with other existing database development tools.

The remainder of this paper is structured as follows:
\Cref{sec:relatedWork} describes related work. 
\Cref{sec:prototype} presents our approach and its prototypical implementation of back-in-time debugging in the database.
\Cref{sec:ttqueries} discusses querying past database states and introduces our extension to SQL.
\Cref{sec:evaluation} evaluates our approach before \cref{sec:conclusion} discusses future work and concludes the paper.


\section{Related Work}
\label{sec:relatedWork}

The first debugger that could reverse execution was EXDAMS for FORTRAN in the late 1960s~\cite{balzer_exdams:_1969}, which stored previous program state on tape.
Later, other reversible debuggers used memory snapshots~\cite{feldman_igor:_1988} or reversible execution~\cite{lieberman1997zstep} to allow stepping backwards through time.

The \emph{omniscient debugger}~\cite{lewis_debugging_2003} was the first debugger to keep the entire program history in memory.
This allows fast jumping between arbitrary points in time and querying the execution history, but creates a large overhead on runtime and memory consumption.
The trace-oriented debugger ~\cite{pothier_scalable_2007} uses a specialized distributed database to better handle large amounts of trace data.
Object  flow  analysis~\cite{lienhard_practical_2008} reduces the required amount of memory by leveraging the VM's garbage collector.
The Path tool suite and its Test-Driven Fault Navigation~\cite{perscheid2013} leverages reproducible test cases in order to distribute the tracing overhead over multiple runs depending on developers' needs.
\emph{SPYDER}~\cite{agrawal_debugging_1993} and the \emph{Slice Navigator}~\cite{treffer2016} are back-in-time debuggers that use dynamic slicing to make following the infection chain more efficient.


\section{Back-in-time Debugging for\texorpdfstring{\\ }{ }Stored Procedures}
\label{sec:prototype}

Back-in-time debuggers typically create a significant overhead on run-time and memory, as they need to trace and record every aspect of an execution.
When debugging programs that run on a database, a back-in-time debugger faces two additional problems:
First, the program's behavior strongly depends on external state.
With live debugging, this sometimes can be mitigated by resetting the database before restarting the program, but a post-mortem analysis can not easily examine intermediate states when they have been changed.
Second, much of the data processing happens outside of the program's scope.
Sometimes, state that has impact on a query's result even remains entirely in the database and is practically invisible to analysis tools.
The only way to examine such state is through specific database queries before it is changed.

Stored procedures add the additional challenge that results of a query can be used in subsequent database statements without having been fetched into the program.
This increases the amount of invisible state.
Furthermore, as trace data usually exceeds program data by orders of magnitude, we expect that tracing the processing of every tuple in a large database is not feasible.

With \tool, we focuses on stored procedures because they are executed close to the data and, thus, more efficient in handling large data sets.
This does not limit the general validity of our results, as a debugger for stored procedures faces all of the challenges described above.

\subsection{Replaying a Stored Procedure}

While using a database to manage application state creates several problems for debugging tools, it also opens up new possibilities to simplify or improve the debugger.

First, the database persists state beyond the program execution.
Even post-mortem, the state can be directly accessed by the debugger.
Second, all instructions handling large data sets (i.e., SQL statements), are declarative.
This means they can be analyzed without knowing how they are physically executed and results are reproducible as long as the underlying data does not change.
Thus, neither do we need to trace the internals of a query's execution, nor do we need to persist the query result.
Third, the database can be used to efficiently manage old state.
In-memory databases with insert-only storage automatically retain old data~\cite{Plattner2009Acd}.
With insert-only, older versions of the database can be easily reproduced.
For other databases, insert-only can be achieved by adding validity columns.
For many business applications, such columns already exist where no data may be deleted for legal reasons.

In conclusion, all that is needed to reproduce the execution of a stored procedure is a sequence of execution steps.
Each step represents an instruction that has side effects on the database or assigns a value or result set to a variable.
As there is, conceptually, no concurrency in a stored procedure, we can use sequential numbering to track the order of steps.
If the step involves a database statement, we also need a timestamp to be able to reproduce the query.

With the recorded trace data, the debugger has enough information available to replay the execution and to re-execute any query.

\subsection{Prototype and Tracing Code Generation}

We developed \tool\ as a prototype implementation of our approach.
Our back-in-time debugger runs in the SAP HANA XS-Engine, a framework for web-applications that is part of the SAP HANA platform.
The user interface is written in HTML5 and JavaScript and queries debugging data via AJAX from the back-end, written in server-side JavaScript.
Traces are stored directly in the database.
Packaged as an XS Application, \tool\ can be directly deployed in a SAP HANA installation.

To be able to trace an execution, we developed a SQLScript pre-processor that parses a stored procedure and adds `INSERT` statements around every instruction to collect the required trace data.
To obtain a trace, the debugger then once has to run the traced code instead of the original procedure.
In addition to the tracing code, the pre-processor also generates SQL functions and views that will be used to obtain variable values at given points in time.
For atomic variables, it is simply a search for its most recent step.

For variables containing result tables, a separate view is generated for each query that assigns to that variable.
Each view contains the original code of its query.
Arguments to the query are initialized using their respective views.
On top of that, a master view is generated that chooses the appropriate query view by looking up the requested step in the recorded trace.

\begin{figure}
	\centering
		\includegraphics[width=\linewidth]{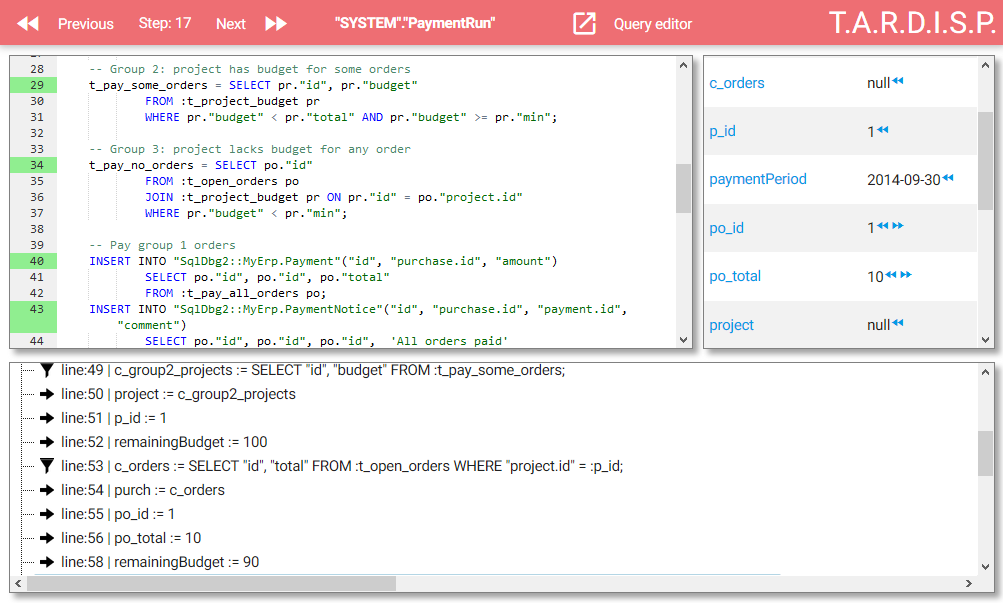}
	\caption{User interface of the \tool\ debugger.}
	\label{fig:odb}
\end{figure}

The user interface looks like a typical debugger and is shown in \cref{fig:odb}.
The top-left part of the screen shows the code.
Above, a toolbar contains buttons allowing developers to step forward and backward.
Between the step buttons, the current step number is shown.
Below, \tool\ shows a tree of all execution steps instead of a stack trace.

Next to the code, current variable values are shown.
Arrow buttons allow to jump to the previous or next assignment of the variable.
For variables containing a query result, the size of the result is shown.
Clicking the variable opens a query window that shows the variable's content and allows the developer to submit time-travel queries, which will be explained in the next section.


\newcommand{\red}[1]{\textcolor{DarkRed}{#1}}
\newcommand{\gr}[1]{\textcolor{Green}{#1}}

\section{\SQLextension}
\label{sec:ttqueries}

In our setup, the debugger has to recreate intermediate results of a stored procedure.
We generalized this feature to enable developers to submit arbitrary queries against the database of any previous point in time.
Furthermore, we allow developers to compare query results from different points in time and even query for changes in the data.
We defined \SQLextension, a super-set of SQL introducing a new keyword and a new operator.

\subsection{A SQL Clause to Query Points in Time}

As an example, we consider a stored procedure that triggers the payment of purchase orders.
Users reported a bug: projects sometimes exceed their budget, which is not supposed to happen.
Developers might start debugging this issue and the related stored procedure with \tool.
However, they soon realize that the number of projects is too large to continue stepping manually.
Instead, they open \tool's SQL console and submit the query shown in \cref{lst:ttravel} to look into the `select_projects` variable, which contains ids, names, and budgets of all projects to be processed.
To better understand the data, they also sum up the total amount of all open purchase orders for each project.

\begin{lstlisting}[language=HanaSQL,float,caption={Example for a time-travel query: select the current total of open orders for previously selected projects at step 1623 of the execution},label=lst:ttravel]
  SELECT pr.id, pr.name, pr.budget, SUM(po.total)
  FROM :selected_projects pr
	JOIN PurchaseOrders po ON po.project_id = pr.id
	WHERE po.status = 'open'
	GROUP BY pr.id, pr.name, pr.budget
	^§AT STEP§ 1623^
\end{lstlisting}

The last line of \cref{lst:ttravel} contains the `^AT STEP^` clause, our extension to SQL that allows developers to explicitly query a specific point in time.
When opening the SQL console, \tool\ automatically provides an `AT STEP` clause with the number of the current debug step, `^1623^` in this case.
Developers can obtain numbers for other steps from the stepping toolbar or the execution tree. 

When the query is submitted, \tool\ applies two transformations before it is passed to the database.
First, all variables are replaced with their corresponding functions or views that were generated during the pre-processing of the stored procedure.
Each function or view receives the specified step number as an argument to be able to produce the correct value or table.
Second, time-stamp filters on the validity columns are added for all tables that are referenced in the query. 
This lets all tables appear exactly as they originally were at the specified point in time.

Now, the query can be submitted to the database and the result is subsequently presented to the user.

\subsection{Time-diff Queries}

Furthermore, \SQLextension\ provides an easier way to find the origin of changes in the data.
To get a better overview about what happened in a segment of code, developers might want to query multiple points in time at once and see the difference in the query result.
In our example, developers are looking for purchase orders that cause projects to exceed their budget.
Using \tool, they identified three important points in time, which we will call \emph{before}, \emph{now}, and \emph{after}.
\emph{Before} is the step where the stored procedure started processing purchase orders, \emph{now} is the current step, and \emph{after} is the final step of the procedure.

In \cref{lst:tdiff}, the query from above was extended to select individual purchase orders for each project.
In line 10, the at-step clause now specifies all three points in time.
This allows developers to query for changes in the data.

Because the steps are named, they can now be used in the query as time qualifiers.
\SQLextension\ introduces the exclamation mark operator that binds an identifier to a step.
In line 6, it is used to limit the search to projects that will exceed their budget before the stored procedure ends;
in line 7, it is used to filter for purchase orders whose status was or will be changed.

\begin{lstlisting}[language=HanaSQL,float=t,caption={Example of a time-diff query: "Select all projects that will go over budget and their respective purchase orders"},label=lst:tdiff]
	SELECT pr.id, pr.name, pr.budget, SUM(po.total), po2.id, po2.status, po2.total
	FROM :selectedProjects pr
	JOIN PurchaseOrders po ON po.project_id = pr.id
	JOIN PurchaseOrders po2 ON po2.project_id = pr.id
	WHERE po.status = 'open'
		AND ^now!^pr.budget > 0 AND ^after!^pr.budget < 0
		AND ^before!^po2.status != ^after!^po2.status
	GROUP BY pr.id, pr.name, pr.budget, 
	         po2.id, po2.status, po2.total
	^§AT STEP§ before=817, now=1623, after=2043^
\end{lstlisting}
\ctable[caption={Result of a time-diff query, with multiple values in some columns. Red indicates a different value at the \emph{before} step, green a different value at the \emph{after} step.},label=tab:diffresult,doinside={\small}]
				{rrrrrrr}{}{%
	\multicolumn{3}{c}{pr.} & \multicolumn{1}{c}{po.} & \multicolumn{3}{c}{po2.} \NN
	\cmidrule(lr){1-3} \cmidrule(lr){4-4} \cmidrule(lr){5-7}
	id & name & budget  & total & id & status & total \ML
	
				&						& \red{1200} & \red{1500} &	  & \red{open} &					\NN
	1			&	Project 1	& 200			 	 & 500 			  & 1 & paid 			 & 1000			\NN
				&						& \gr{-300}  & \gr{0}		  &	  &						 &					\ML

				&						& \red{1200} & \red{1500} &	  & 					 &					\NN
	1			&	Project 1	& 200				 & 500 			  & 2 & open 			 & 500			\NN
				&						& \gr{-300}  & \gr{0}		  &	  & \gr{paid}	 &					\ML
}

\Cref{tab:diffresult} shows a possible result for this query, with one project that goes over budget and two associated purchases.
The values for budget and total of outstanding payments are shown for each of the three steps.
Of the two purchase orders, the first was already marked as "paid" before the current step, the other is about to be paid.

Analyzing this data, it becomes clear that the payment of the second order is what causes the project budget to become negative.
Developers can now click on the green "paid" value to jump exactly to the point in time where this order is updated, knowing that at this point they are very likely to be close to the defect they are looking for.

To produce this result, the query has to be executed three times, once for each point in time, without the time-specific where-conditions.
The partial results are outer-joined on the primary key attributes, then the time-specific filter conditions are applied.
\tool\ rewrites the query to fit everything into a single select statement, which allows the database to compute the entire result in one request.

In the user interface, changes in the values are highlighted in as shown in \cref{tab:diffresult}.
If a value changed since the \emph{before} step, the old value is shown in red; if a value will change until the \emph{after} step, the new value is shown in green.
When possible, the tuple creation timestamps are used to allow developers to jump exactly to the point in time where the change occurred.


\section{Evaluation}
\label{sec:evaluation}

\ctable[caption={Average execution time for running a stored procedure without and with tracing and for reproducing its result.},label=tab:measure1,doinside={\small},width=\linewidth]
				{lc>{\raggedleft \arraybackslash}p{1.3cm}RR}{}{
				&		& Normal run& Run with tracing  & Reproducing the result \ML
	
			 &	(a) &	1.27 s		& 1.35 s		& 1.21 s		\NN
Proc. 1& (b)	&	1.61 s		& 1.74 s		& 1.54 s		\NN
			 &	(c) &	1.73 s		& 1.85 s		& 1.60 s		\ML

			 &	(a) &	1.24 s		& 1.32 s		& 1.19 s		\NN
Proc. 2& (b)	&	1.61 s		& 1.72 s		& 1.55 s		\NN
			 &	(c) &	1.69 s		& 1.80 s		& 1.63 s		\ML
}

\ctable[caption={Average execution times for queries executed normally or as time-diff query.},label=tab:measure2,doinside={\small},width=\linewidth]
				{lcRR}{}{
				&		& Regular query& Time-diff query \ML
	
			 &	(a) &	1.21 s		& 1.51 s	\NN
Query 1& (b)	&	1.74 s		& 2.12 s	\NN
			 &	(c) &	1.79 s		& 2.23 s	\ML

			 &	(a) &	61 ms		&  75 ms	\NN
Query 2& (b)	&	81 ms		& 100 ms	\NN
			 &	(c) &	88 ms		& 109 ms	\ML
}

We evaluated our \tool\,debugger on a real-world SAP project that has been developed with one of the largest European retail companies. 
This project is called \emph{Point of Sales Explorer}~\cite{plattner2015memory} and allows category managers to see a collection of the most important key performance indicators (KPIs) for several thousands of products in a unified dashboard.
Based on more than 2 billion records of point of sales data, this application aggregates on the fly the requested KPIs with the help of SAP HANA and allows users to further refine them by stores, vendors, or products as they like. 
In order to implement flexible requests such as returning the revenue and margin per week of current and last year, this application makes heavy use of SQLScript. 
For that reason, it is a proper candidate to measure performance and interview its developers with respect to our \tool\ tool. 

\subsection{Performance Measurements}

To evaluate the performance of our approach, we deployed \tool\ on a copy of the productive system.
We selected two stored procedures from the Point of Sales Explorer and compared the runtime with and without tracing.
We ran each procedure with different arguments that would lead to intermediate results of different sizes.
As \cref{tab:measure1} shows, we found the overhead of tracing to be consistently between six and seven percent.

We also measured the time it takes to reproduce the results of each stored procedure.
The measurements in \cref{tab:measure1} show that reproducing the result is slightly faster than executing the stored procedure.
This was expected because for variables containing atomic values the value could be taken directly from the trace.
Only for variables containing table results, database queries had to be executed again.

Finally, we selected two queries from the stored procedures as examples a developer might submit through the debugger's SQL console and measured their execution time both as a regular query and as a time-diff query.
Again, we used different filter arguments to produce different result sizes.
As \cref{tab:measure2} shows, in each case the runtime overhead of time-diff queries is around 25 percent.

Developers will avoid tools that are too slow to allow fluent working because every delay distracts from the problem to be solved. 
Back-in-time debuggers often suffer from this problem.
Our measurements show that the overhead created by \tool\ is small enough to make it a feasible alternative to existing debugging tools.

\subsection{Interviews}

Besides measuring performance, we also demonstrated our back-in-time debugger to two backend developers of the Point of Sales Explorer. 
These persons have written most of the SQLScript and so know all the challenges when developing close to the database. 
Both had the chance to apply \tool\ in their daily work while we provided tool support and observed them in the background. 
Finally, we have conducted an in-depth interview in order to receive feedback and further improve our tool.

We found out that the main reason for writing SQLScript is to positively influence the optimizer of SAP HANA in order to speed up the overall user request.
If something went wrong at the database-level, the existing tool support was not sufficient. 
The interviewed developers often had to guess why a specific sub-query was not return the expected results. 
For this, they liked \tool\ as it could not only show formerly hidden results but also helped them to understand and follow back the reasons behind a failure cause. 
The greatest improvement was reported for debugging stored procedures that take a table of data as an argument, whose cumbersome preparation has to be done only once, using \tool.
Besides that, both developers wished for some new features such as the ability to define temporary tables for testing code and queries.
With this feedback, we see a strong need for better debugging support at the database-level and are eager to further improve our approach.


\section{Conclusion}
\label{sec:conclusion}

\balance
We presented an approach for bringing back-in-time debugging down to SQL and stored procedures.
Our \tool\ debugger is an implementation of our approach that runs on the SAP HANA in-memory database.
With \tool, developers can step forward and backward through the execution of a stored procedure.
We introduced \SQLextension, a super-set of SQL, that allows developers to query the database from any point in the execution, compare the data of multiple points in time with one query, and even query for changes in the data.

Our evaluation has shown that the runtime overhead caused by execution tracing is small enough to let \tool\ be the default choice for debugging stored procedures.
Interviews with developers have confirmed the usefulness of back-in-time debugging for finding root causes in database programs.

Future work will move in two directions.
First, we shall bring more advanced debugging techniques that use a combination of trace data and code analysis to help identify failure causes, such as dynamic slicing~\cite{agrawal_dynamic_1990} down to the database.
Second, while developers now have the tools to efficiently debug each layer of a complex application, the tool support to track infection chains across layers and system boundaries is still lacking.
We want to develop a back-in-time debugger that will allow developers to debug seamlessly across application layers.

%
\bibliographystyle{abbrv}
\bibliography{sigproc}  

\end{document}